# Kinetic Solvers with Adaptive Mesh in Phase Space for Low-Temperature Plasmas

**Vladimir Kolobov,** [a,b,1] **Robert Arslanbekov** [a] **and Dmitry Levko** [a]

[a] CFD Research Corporation, Huntsville, AL 35806, USA
[b] The University of Alabama in Huntsville, AL 35899, USA

E-mail: vladimir.kolobov@cfdrc.com

**Abstract**. We describe the implementation of 1d1v and 1d2v Vlasov and Fokker-Planck kinetic solvers with adaptive mesh refinement in phase space (AMPS) and coupling these kinetic solvers to Poisson equation solver for electric field. We demonstrate that coupling AMPS kinetic and electrostatic solvers can be done efficiently without splitting phase-space transport. We show that Eulerian fluid and kinetic solvers with dynamically adaptive Cartesian mesh can be used for simulations of collisionless plasma expansion into vacuum. The Vlasov-Fokker-Planck solver is demonstrated for the analysis of electron acceleration and scattering as well as the generation of runaway electrons in spatially inhomogeneous electric fields.

## 1. Introduction

Simulations of low-temperature plasma require coupling kinetic solvers with electromagnetic solvers. With increasing computing power, mesh-based kinetic solvers become competitive to the particle-based solvers.[1,2] Using adaptive mesh is a proven method for reducing computational cost of both mesh-based kinetic solvers and electromagnetic solvers. The challenge appears when coupling solvers of different dimensionality. While kinetic solvers operate in phase space, electromagnetic solvers operate in configuration (physical) space.

We have previously shown that splitting velocity and configuration spaces is one way to resolve the coupling challenge for kinetic solvers with adaptive mesh in phase space (AMPS).[3] Here, we further explore this challenge for problems of lower dimensionality. Using symmetry is a well-known method to reduce dimensionality of kinetic solvers. In the present paper, we describe the implementation and application of 1d1v and 1d2v kinetic solvers with AMPS and the coupling of these solvers to 1d Poisson solver for electric field. The choice of coordinate system for kinetic equations depends on the problem type. The best coordinate system in physical space is dictated by the symmetry of the problem. The best coordinate system in velocity space depends on relative importance of collisions and forces. Usually, collision operator is close to diagonal in spherical coordinate system while the force term is nearly diagonal in cylindrical coordinate system.

Collisions of charged particles with neutrals are described using spherical coordinates in velocity space. For transport in configuration (physical) space, we use Cartesian, cylindrical or spherical coordinate systems depending on problem type. Collisions are divided into two types. Boltzmann-type collisions are associated with a large change of particle momentum – they are described by an integral operator in velocity space. Fokker-Planck-type collisions are associated with small changes of momentum. They are described by Fokker-Planck-type differential operators in velocity space. We solve Vlasov and Fokker-Planck kinetic equation using Finite Volume method with octree Cartesian mesh, without splitting physical and velocity spaces. As a result, mesh adaptation in velocity space triggers mesh adaptation in physical space. This creates some difficulties for calculation of particle density and coupling kinetic solvers to Poisson solver for calculation of electrostatic field.

---

[1] To whom any correspondence should be addressed.

In this paper, we describe the implementation and application of 1d1v and 1d2v Vlasov and Fokker-Planck kinetic solvers and the coupling of these solvers to the Poisson solver for electric field. We demonstrate that coupling Vlasov and Poisson solvers can be done efficiently without splitting phase-space transport. We also demonstrate that (Eulerian) fluid and kinetic solvers with dynamically adaptive Cartesian mesh can be used for simulations of multi-dimensional problems of collisionless plasma expansion into vacuum.

Collision processes in the Fokker-Planck solver have been implemented using spherical coordinates in velocity space for simulations of electron kinetics in weakly-ionized plasma. We demonstrate these solvers for analysis of electron acceleration by electric fields and scattering by neutral atoms as well as electron impact ionization of atoms and generation of runaway electrons in spatially inhomogeneous electric fields.

## 2. Coupling Kinetics with Electrostatics

A whole spectrum of mesh-based kinetic solvers has been explored ranging from those without splitting 6d phase space transport[4] to those splitting phase space transport into 1d advection for each dimension.[5] To explore best strategy of coupling AMPS kinetic solvers with electrostatic solvers, we study here 1d1v and 1d2v problems using Basilisk - an open source framework for solving partial differential equations on adaptive Cartesian meshes.[6]

When using adaptive mesh in phase space without splitting velocity and physical spaces, calculations of VDF moments (such as particle density and fluxes) by integration of the VDF over velocity space becomes challenging. The problem of coupling AMPS Vlasov and Poisson solvers has been clearly demonstrated in Ref. [7]. To overcome this challenge, we have developed a methodology illustrated in Figure 1 for the quad-tree grid.

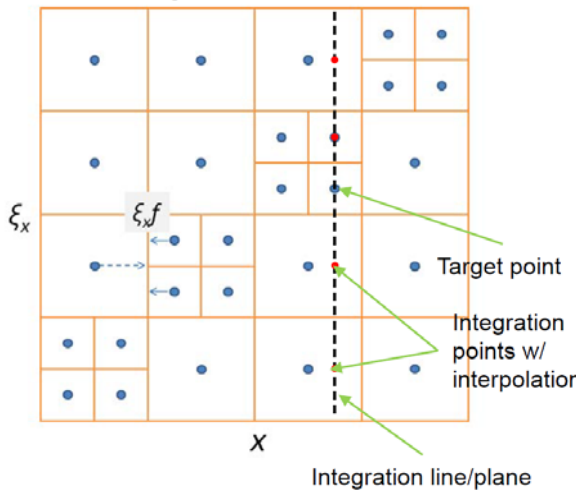 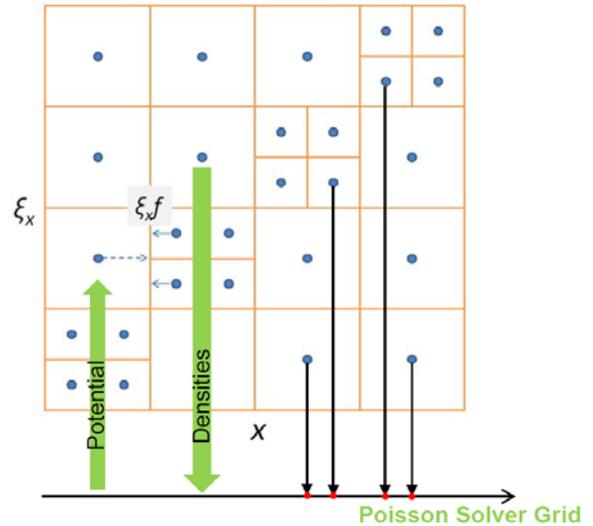

*Figure 1. Calculation of velocity space integrals (left) and coupling 1d1v Vlasov and 1d Poisson solvers (right).*

To compute VDF moments (density, mean velocity, temperature, etc.) an integration along an x = const line must be performed. In our method, all cells containing this line are identified, and 2$^{nd}$ order interpolation (involving all neighboring cells) is carried out to compute the VDF required for the integration to proceed.

A one-dimensional Poisson solver is coded inside the 1d1v Vlasov solver. For coupling Vlasov and Poisson solvers, the 1d1v quadtree phase-space grid is projected into 1d Poisson-solver grid using cell centered x-locations. The calculated particle densities are then transferred from the 1d1v grid to the 1d grid. After solving the Poisson equation on the 1d grid, electric potential and electric field are returned back at cell centers and faces of the 1d1v phase-space grid. This method is also used for the 1d2v octree phase-space grid. Below, we demonstrate some results obtained with the coupled solvers.

### 3. Collisionless Plasma Expansion

Plasma expansion into vacuum is a fundamental problem with important applications in space science, laser ablation, electric propulsion and arc interactions with electrodes. The problem is associated with large variations of plasma density within a computational domain calling for a hybrid fluid-kinetic solver.[8] In this section, we describe our recent advances of the fluid and kinetic solvers for collisionless plasma.

*3.1. Fluid Model for Cold Ions*

The simplest fluid model of collisionless plasma with cold-ions and Boltzmann electrons comprises of a set of equations for the ion and electron densities, $n_i$ and $n_e$, ion mean velocity $\vec{u}_i$, and the electrostatic potential, $\varphi$:

$$\frac{\partial n_i}{\partial t} + \nabla \cdot (n_i \vec{u}_i) = 0, \tag{1}$$

$$m_i \frac{\partial \vec{u}_i}{\partial t} + m_i (\vec{u}_i \cdot \nabla) \vec{u}_i = -e\nabla\varphi, \tag{2}$$

$$\nabla^2 \varphi = \frac{e}{\epsilon_0}(n_e - n_i), \tag{3}$$

The electron density is described by the Maxwell-Boltzmann relationship:

$$n_e = n_0 \exp\left(\frac{e\varphi}{k_B T_e}\right), \tag{4}$$

with (constant in space and time) electron temperature, $T_e$.

We have developed a second-order scheme to solve the set of Eqs. (1-4) on dynamically adaptive Cartesian mesh using the approach proposed in Ref. [9]. The non-linear Poisson-Boltzmann equation is solved by converting it to a Poisson–Helmholtz equation. The linear term is treated implicitly while the remaining non-linear term is treated explicitly executing non-linear iterations until a given tolerance is reached. For the linear solver, we use an efficient multigrid solver. A typical number of iterations of the Poisson-Helmholtz solver is 2–5 to achieve convergence within a tolerance of 10-8–10-6 and a typical number of non-linear (Newton) iterations is 5–10 for a tolerance of 0.01–0.1, which were sufficient to achieve converged results.

Figure 2 shows results of our calculations for a planar plasma expansion. This 1d problem was studied in Ref. [10] using both Eulerian and Lagrangian approaches with static mesh. Our results show good agreement with the results of [10] for both spatial and temporal profiles. In particular, very sharp ion density peaks are predicted with peak widths of a few $\lambda_D$.

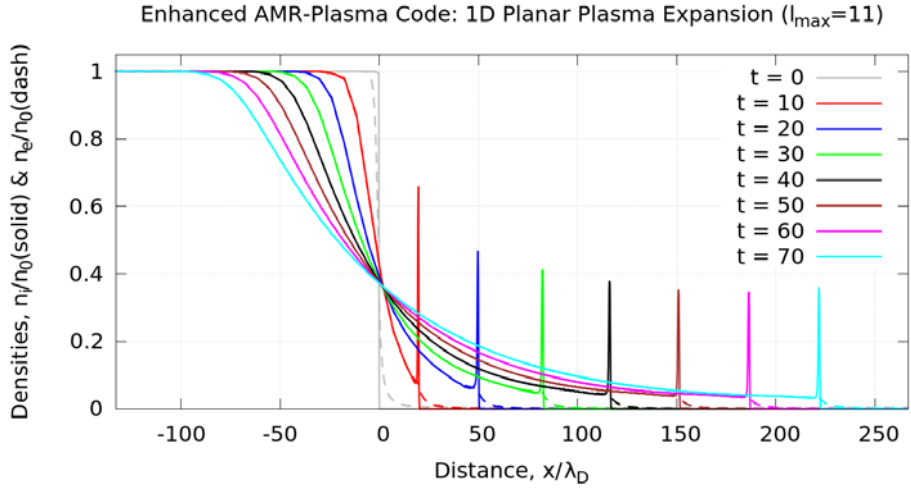

Figure 2. Calculated electron and ion densities for one-dimensional plasma expansion.

Following Ref. [10], we studied effects of initial ion density profile on plasma dynamics. The initial profile was selected in the form $n_i(t=0, x) = 0.5 - \arctan(\pi x/l)/\pi$, with the widths of $l = 1, 5,$ and 20. According to Ref. [10], the maximum electric field occurs behind the outermost ions, at the point where the electron and ion densities are equal. Positive ions at that position tend to catch up with those ahead, producing a peak in the ion density. Our results are in excellent agreement with these theoretical considerations and numerical results.

We have further extended our fluid solver for multi-dimensional problems. Figure 3 shows an example of 3d simulation using octree adaptive mesh. The AMR capabilities allow resolving the narrow space-charge zone moving with 3-4x ion-sound speeds while keeping the rest of computational domain (~400 λD) coarse for efficient computations and fast convergence rates. The expansion zone remained closely hemi-spherical up to several 100's λD away from the boundary.

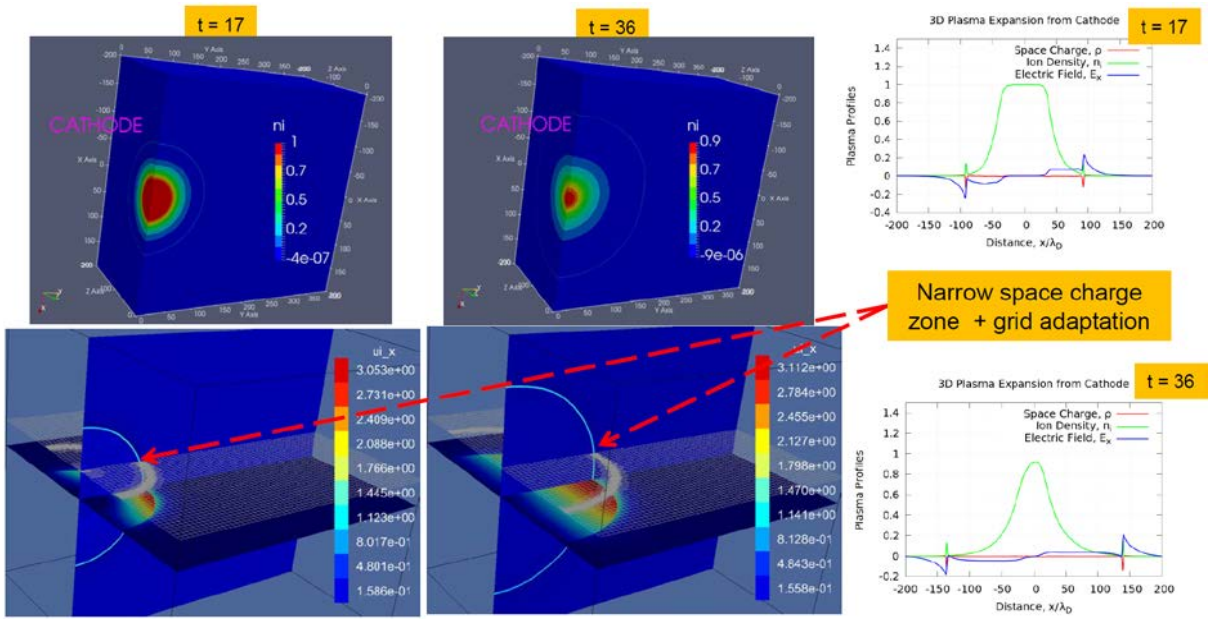

*Figure 3. Simulations of 3D plasma expansion from a cathode.*

## 3.2. Vlasov-Poisson Solver

We have developed a Vlasov solver for ions using quadtree Cartesian mesh in 1d1v phase space. Plasma expansion with the coupled Vlasov-Poisson solver was studied for the planar case. Figure 4 shows spatial distributions of electron and ion densities and the electric field for the case of ion/electron temperature ratio of 0.1, with initial ion density profile as a step function. It is seen that the electric field has a peak at the ion front however no peaks of the ion density are observed. These peaks in the fluid model are artifacts of the numerical treatment of the non-linear term in the ion momentum equation ("wave-breaking" mechanism discussed in Ref [10]). Such a term is absent in the Vlasov formulation.

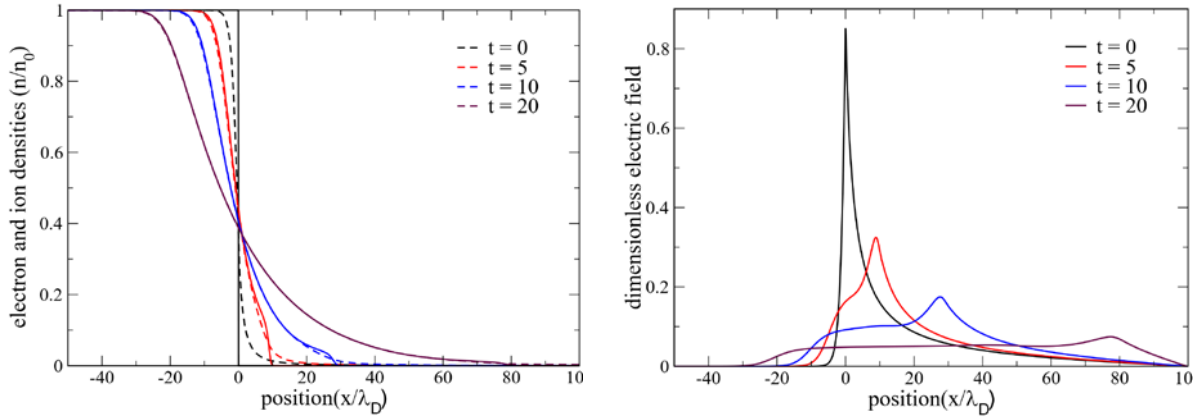

*Figure 4. Dynamics of planar plasma expansion with a step initial ion density profile.*

For a smooth initial ion density profile with a width about one Debye length, we obtained peaks of the ion density profiles at times > 5 in accordance with the fluid models for smooth initial ion density profiles. However, these peaks disappeared at later times (see Figure 5).

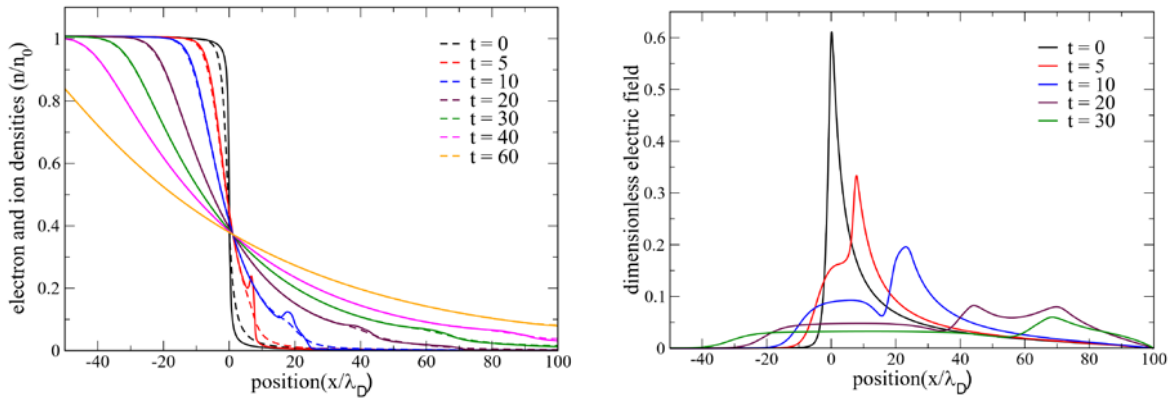

*Figure 5. Dynamics of planar plasma expansion with a smooth initial ion density profile.*

The mean ion velocities in our simulations reach values of 4-5 in accordance with other (Lagrangian) results (Figure 6). Velocities beyond the ion front depend strongly on the physical model (e.g., fluid vs kinetic) and on the numerical scheme (order of spatial and time accuracy, damping, etc.).

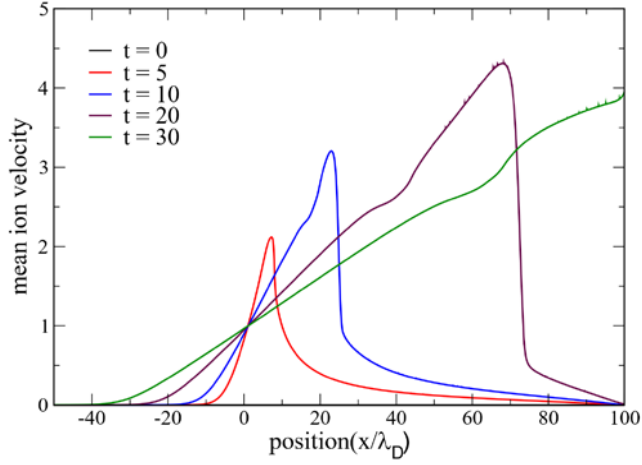

*Figure 6. Mean ion velocity for planar plasma expansion with a smooth initial ion density profile.*

In conclusion to this section we can say that the Vlasov-Poisson solver with AMR can be used for simulations of collisionless plasma expansion provided that adequate grids are used in both physical and velocity spaces. The Eulerian fluid code for cold ions and AMR can also be used well to study plasma expansion dynamics in multi-dimensions. We plan to combine these codes into a hybrid kinetic-fluid framework to study collisionless plasma expansion and other problems that can be described by this model.

## 4. Fokker-Planck Kinetic Solver for Electrons

The Fokker-Planck kinetic equation for the velocity distribution function of charged particles in non-magnetized plasma can be written in the form:

$$\frac{\partial f}{\partial t} + \nabla \cdot (\mathbf{v}f) + \frac{1}{m}\frac{\partial}{\partial \mathbf{v}}\left[\left(F(\mathrm{v})\frac{\mathbf{v}}{\mathrm{v}} - e\mathbf{E}\right)f\right] + \frac{\partial}{\partial \mathrm{v}_i}\left[D_{ij}\frac{\partial f}{\partial \mathrm{v}_i}\right] = I \quad (5)$$

The left side of Eq. (5) describes particle acceleration by the electric force and collision processes with small changes of energy and momentum, which are written in the Fokker-Planck form. The friction force, $F(\mathrm{v})\mathbf{v}/\mathrm{v}$, due to energy loss in collisions is directed towards zero velocity. Small-angle scattering is described by a diffusion tensor

$$D_{ij} = \nu \mathrm{v}^2 \left(\delta_{ij} - \frac{\mathrm{v}_i \mathrm{v}_j}{\mathrm{v}^2}\right) \quad (6)$$

where $\nu(\mathrm{v})$ is the collision frequency. Quasi-linear model of wave-particle interactions also results in a Fokker-Planck form.

### 4.1. Spherical Coordinates in Velocity Space

Let us consider a one-dimensional planar problem and use spherical coordinates in velocity space with the axis directed along the electric force. Assuming that the distribution function does not depend on the azimuthal angle, we can rewrite Eq. (1) in the form:

$$\frac{\partial f}{\partial t} + v\mu \frac{\partial f}{\partial x} - \frac{eE}{m}\left(\mu \frac{\partial f}{\partial v} + \frac{1-\mu^2}{v}\frac{\partial f}{\partial \mu}\right) = \frac{1}{v^2}\frac{\partial}{\partial v}(v^2 F f) + D_\mu \frac{\partial}{\partial \mu}\left[(1-\mu^2)\frac{\partial f}{\partial \mu}\right] + I \quad (7)$$

where $\mu = \cos\theta$. Now, the diffusion tensor (6) has only one component, $D_\mu(v)$. Introducing $Y = v^2 f$, Eq. (7) can be written in the conservative form:

$$\frac{\partial Y}{\partial t} + \frac{\partial}{\partial x}(UY) + \frac{\partial}{\partial v}(VY) + \frac{\partial}{\partial \mu}[WY] = \frac{\partial}{\partial \mu}\left[D_\mu \frac{\partial Y}{\partial \mu}\right] + v^2 I \quad (8)$$

where

$$U = \mu v; \qquad V = eE\mu - F; \qquad W = eE\frac{(1-\mu^2)}{mv}.$$

In the absence of scattering and ionization, a semi-analytical solution of Eq. (8) can be obtained by integration along characteristics:

$$\frac{dv}{dx} = \frac{eE\mu - F(v)}{mv\mu}; \qquad \frac{d\mu}{dx} = \frac{eE}{m}\frac{(1-\mu^2)}{v^2\mu}; \qquad (9)$$

The boundary condition $f(x=0) = \Phi(v_0, \mu_0)$ defines the solution $f(x,v,\mu) = \Phi(v_0(v,\mu), \mu_0(v,\mu))$ where $\mu_0$ and $v_0$ denote the values of $\mu$ and $v$ at $x=0$. In the absence of energy losses ($F(v)=0$), Eqs. (9) can be integrated to give:

$$\frac{1-\mu^2}{1-\mu_0^2} = \frac{w_0}{w_0 - e\varphi(x)} \qquad (10)$$

where $w_0 = mv_0^2/2$ is the electron kinetic energy at $x=0$, and $\varphi(x) = -dE/dx$ is the electrostatic potential. Eq. (10) illustrates how rapidly the electron trajectories (characteristics) converge to $\mu = 1$ due to electron acceleration by the electric field.

### 4.2. Fokker-Planck Solver for Electrons

The FP equation (8) was solved numerically using adaptive Cartesian mesh in 1d2v phase space $(x, v, \mu)$. Figure 6 illustrates the FP solver for the problem of electron acceleration and scattering in a spatially inhomogeneous electric field. It is assumed that the electric field, $E(x)$, decreases linearly at $x < d$, and is zero at $d < x < L$. Electrons are injected at $x = 0$ with a Maxwellian distribution and are absorbed at the right boundary, $x = L$. In the ($v_\parallel, v_\perp$) plane, the VDFs form rings with increasing radius, which are gradually filled by particles (colour) due to scattering.

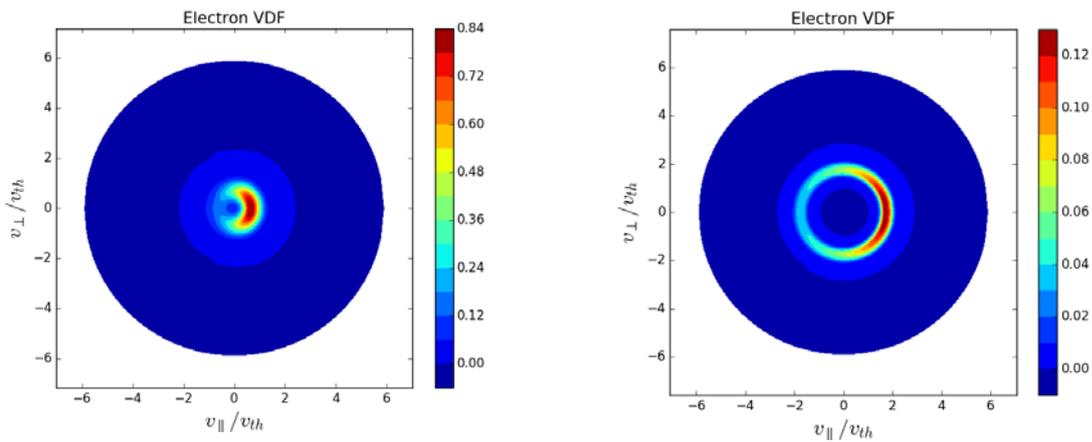

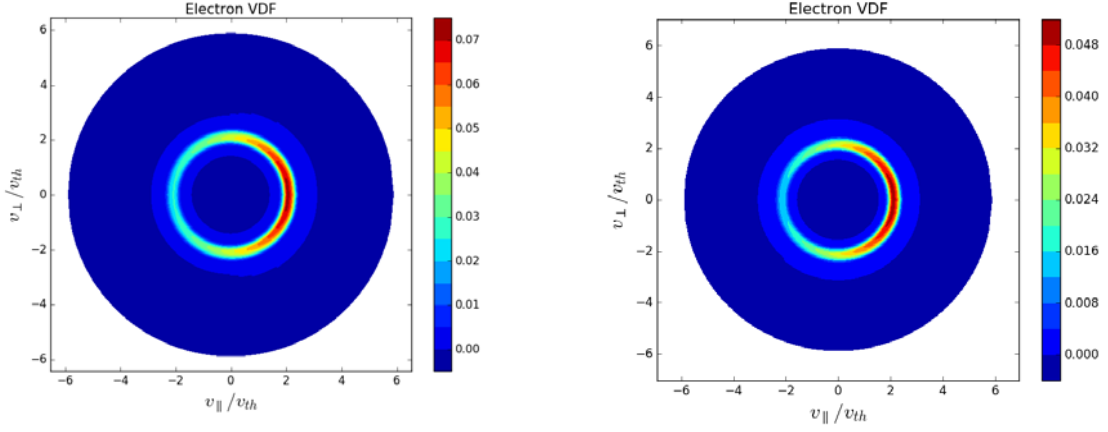

*Figure 7. Electron VDF calculated with FP solver at x/L = 0.1, 02, 0.4, and 0.8, for d/L = 0.4.*

*4.3. Runaway Electrons*

Electron runaway in spatially inhomogeneous electric field is a complicated process involving acceleration, energy loss in collisions, scattering and ionization.[11] The energy loss function $F(w)$ has a maximum at about 100 eV, and scattering changes from nearly-isotropic to forward-directed with increasing electron energy. To apply our FP solver for analysis of electron runaway the ionization term was approximated in the simplest form[11]

$$I = \delta(\mathbf{v}) \int \frac{F(v)}{\varepsilon_0} f d\mathbf{v} \qquad (11)$$

where $\delta(\mathbf{v})$ is a velocity distribution of the secondary electrons, and $\varepsilon_0$ is the cost of ionization. As in the previous section, we consider a linearly decreasing electric field. For easier interpretation of the obtained results, we assume $F(v) = const = F_0$ and $eE_0/F_0 = 5$, which guaranties the generation of runaway electrons at $eE(x)/F_0 > 1$.

Here, we consider a 1d1v case using the forward-backward scattering model.[12][13] The energy-dependence of the scattering rate is approximated in the form $\exp(-0.7v^2)$. Figure 8 illustrates the formation of VDF under effects of acceleration, energy loss, and scattering. Electrons are injected at the left boundary ($x = 0$) with a Maxwellian distribution, move along characteristics (stream lines in the left part of Figure 8), and get absorbed at the anode, at $x = L$. Elastic scattering produces jumps between positive and negative values of v. Due to the rapid decrease of the scattering rate with electron energy there is asymmetry between positive and negative v: slow electrons have near-isotropic VDF, whereas runaway electrons have strongly anisotropic VDF (see the right part of Figure 8).

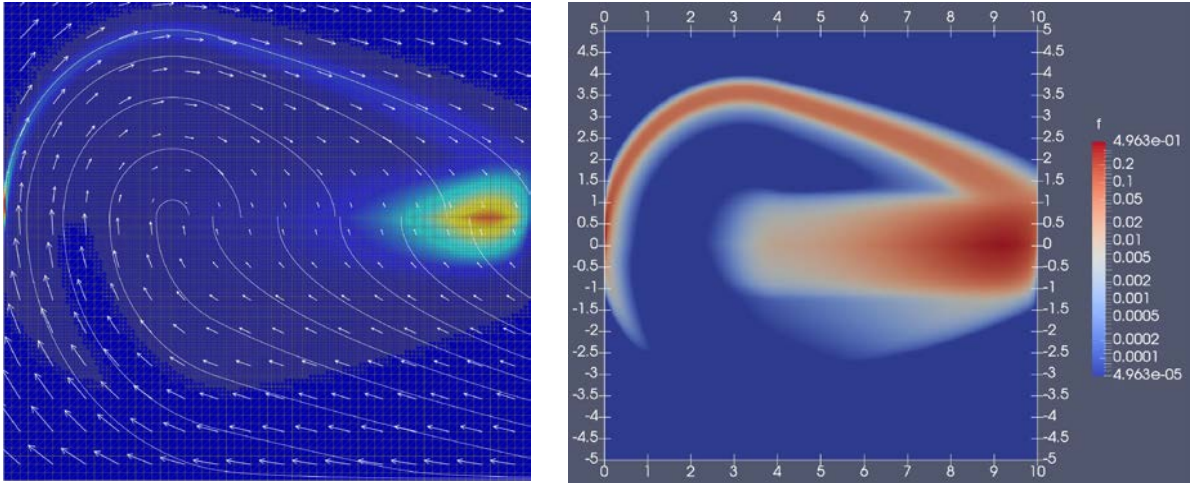

*Figure 8. Streamlines (characteristics) and adapted computational mesh in 1d1v phase space (left). The calculated VDF (colour map in log scale) in the 1d1v phase space. No ionization.*

Finally, we include the generation of secondary electrons by electron-impact ionization of the background gas. The velocity distribution of the secondary electrons is assumed in the form $\delta(v) = \sqrt{\xi/\pi} \exp(-\xi v^2)$, which at $\xi \to \infty$ becomes a delta-function considered in the analytical model.[11] Figure 9 shows the calculated VDF and spatial distributions of the ionization rate, electron density and temperature. Note the difference in the absolute values of the VDF compared to Figure 8, which is due to the large number of the secondary electrons generated in the present case. The axial distributions of the plasma parameters shown on the right part of Figure 9 correspond to the typical distributions observed in the cathode region of short glow discharges without positive column (see Ref. [11] for details).

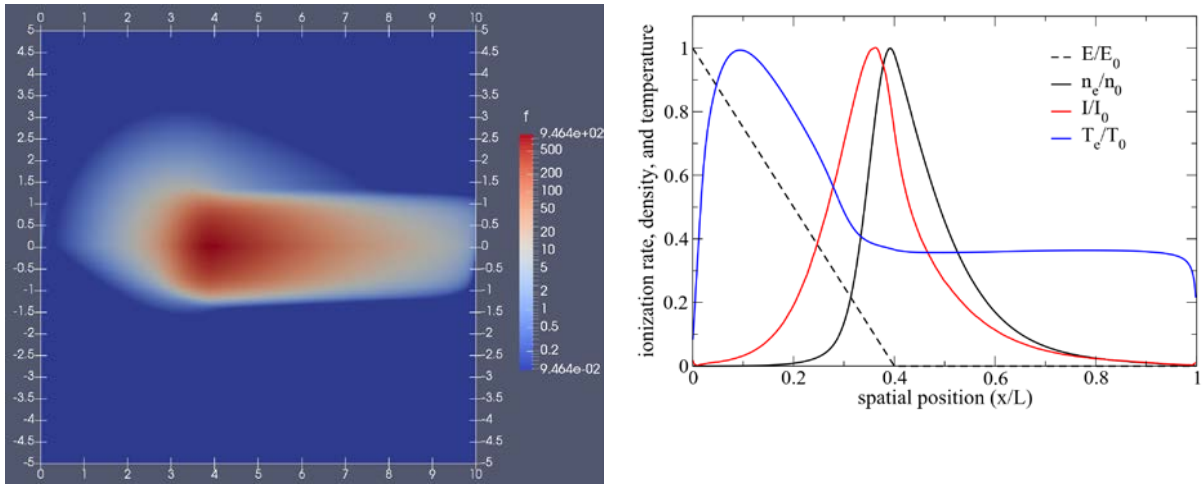

*Figure 9. The calculated VDF (left, colour map in log scale) in the 1d1v phase space (left) and spatial distributions of normalized macro-parameters (right) with ionization.*

### 5. Conclusions

We have described the implementation and application of 1d1v and 1d2v Vlasov and Fokker-Planck kinetic solvers with adaptive mesh in phase space (AMPS). We have demonstrated that coupling kinetic solvers with electrostatic (Poisson) solver can be done efficiently without splitting phase-space transport. We have shown that Eulerian fluid and kinetic solvers with dynamically adaptive Cartesian mesh can be used for simulations of collisionless plasma expansion.

We have implemented 1d2v Fokker-Planck solver using AMPS with spherical coordinates in velocity space for electron kinetics in partially ionized collisional plasma. We have demonstrated these solvers for analysis of electron acceleration, scattering, and generation of runaway electrons in spatially inhomogeneous electric fields. This work opens new opportunities for solving plasma problems that can be described by Vlasov and Fokker-Planck equations in 1d2v phase space.

**Acknowledgments**

This work was partially supported by the DOE SBIR Project DE-SC0015746, by the US Department of Energy Office of Fusion Energy Science Contract DE-SC0001939, and by the NSF EPSCoR project OIA-1655280 "Connecting the Plasma Universe to Plasma Technology in AL: The Science and Technology of Low-Temperature Plasma".